\documentclass[10pt,conference]{IEEEtran}

\usepackage{quantikz}
\usepackage[english]{babel}
\usepackage[utf8]{inputenc}
\usepackage[colorinlistoftodos, color=green!40, prependcaption]{todonotes}
\usepackage{amsthm}
\usepackage{mathtools}
\usepackage{physics}
\usepackage{braket}
\usepackage{listings}

\usepackage{graphicx}
\usepackage{adjustbox}
\usepackage{placeins}
\usepackage[T1]{fontenc}
\usepackage{lipsum}
\usepackage{csquotes}
\usepackage{array}
\usepackage[pdftex, pdftitle={Article}, pdfauthor={Author}]{hyperref} 
\usepackage{booktabs}
\usepackage{amssymb}
\usepackage{algorithmic}
\usepackage{bm}
\usepackage[ruled,vlined]{algorithm2e}
\usepackage{placeins}
\usepackage[caption=false]{subfig}

\usepackage{float}
\usepackage{dcolumn}
\usepackage{bm}

\usepackage{diagbox}
\usepackage{multirow}
\newcolumntype{?}{!{\vrule width 1pt}}
\usepackage{makecell}

\begin{document}

\title{Leveraging Correlated Decoding for Bias-Tailored Compass Codes}
\author{\IEEEauthorblockN{Arianna Meinking}
\IEEEauthorblockA{\textit{Dept. of Physics} \\
\textit{Duke University} \\
 Durham, NC \\
arianna.meinking@duke.edu}
\and
\IEEEauthorblockN{Julie A. Campos}
\IEEEauthorblockA{\textit{Dept. of Physics} \\
\textit{Duke University} \\
 Durham, NC \\
julie.campos@duke.edu}
\and
\IEEEauthorblockN{Kenneth R. Brown}
\IEEEauthorblockA{\textit{Dept. of Elec. Comp. Eng} \\
\textit{Dept. of Physics} \\
\textit{Dept. of Chemistry} \\
\textit{Duke University} \\
 Durham, NC \\
kenneth.r.brown@duke.edu}
}

\maketitle

\begin{abstract}
Quantum error correction (QEC) is often implemented on hardware that experiences biased noise, where dephasing errors occur more frequently than other errors. This has motivated many recent efforts to develop bias-tailored QEC codes, such as the Clifford-deformed compass codes: a family of codes that achieve high thresholds under biased dephasing noise. We perform circuit-level simulations of the Clifford-deformed elongated compass codes under a biased noise model and evaluate code thresholds using standard minimum weight perfect matching (MWPM) and correlated MWPM. We find that correlated decoding enhances thresholds for all noise biases relative to standard MWPM under circuit-level noise. Our results demonstrate that correlated decoding leads to a higher relative gain in thresholds compared to standard MWPM when applied to codes with asymmetric stabilizers under biased noise.
\end{abstract}


\section{Introduction}\label{sec:intro}
Quantum error correction (QEC) is necessary for fault tolerant quantum computation \cite{aharonov1997fault,knill1998resilient,gottesman1997stabilizer,nielsen2001quantum}. For QEC to be most effective, we should co-design QEC codes and decoders with the noise models of quantum computing architectures \cite{dennis2002topological, wang2003confinement, calderbank1996good, knill1998resilient, aharonov1997fault}. For example, trapped ions, superconducting qubits, and neutral atoms can exhibit strongly asymmetric noise in which there are more frequent Pauli-$Z$ errors than Pauli-$X$ errors \cite{aliferis2009fault,lescanne2020exponential,grimm2020stabilization,berdou2023one,bocquet2024quantum,cong2022hardware_neutralatoms,wu2022erasure,kang2023quantum,kubica2023erasure,teoh2022dual}, motivating bias-tailored QEC \cite{tuckett2018ultrahigh,tuckett2019tailoring,stephens2013high,Tuckett2020FTSCBias,Tuckett2020FTSCBias,Martinez2026leveragingbiasatcircuitlevel,bonilla2021xzzx,li20192d,dua2022clifford,setiawan2024tailoring,claes2023tailored,sahay2023tailoring,san2023cellular, Khosravani2026heterogeneousQECcodes}.

Two types of bias-tailored QEC codes include Clifford-deformed surface codes \cite{bonilla2021xzzx, tuckett2018ultrahigh,tuckett2019tailoring,Tuckett2020FTSCBias, dua2022clifford} and elongated compass codes \cite{li20192d,huang2020fault}. Both of these code families demonstrate improved performance under biased Pauli noise compared to symmetric depolarizing noise. The $XZZX$ surface code famously produces a threshold of $50\%$ under infinite bias towards any single Pauli error under code-capacity noise \cite{bonilla2021xzzx}. This improvement in threshold can be attributed to the symmetries imposed on the code by the Clifford deformation. The elongated compass codes are tailored for moderate noise bias towards Pauli-$Z$ errors by fixing gauge operators on the 2D compass codes according to the elongation parameter $\ell$ \cite{li20192d,bombin2015gauge,huang2020fault, paetznick2013universal}. The higher the elongation parameter $\ell$, the better the code can detect and correct Pauli-$Z$ errors \cite{li20192d}. In previous work, we combined single-qubit Clifford deformations with elongated compass codes to produce a class of codes we call the Clifford-deformed elongated compass codes \cite{campos2025}. These codes have high thresholds for moderate and high biases under code-capacity and phenomenological noise. In this work, we extend these studies to the circuit-level and consider improved decoders. 

The minimum-weight perfect matching (MWPM) \cite{edmonds1965paths} decoder typically used on these codes is not optimal. It requires a simplification of the matching graph to avoid hyperedges (edges between more than two detectors). These hyperedges contain information about correlated errors. One of the best examples of these correlations is between $X$ and $Z$ errors due to $Y$ errors. When there are few hyperedges in the matching graph, MWPM performs well \cite{higgott2022pymatching,Gidney2021stimfaststabilizer,tomita2014low,Higgott2023FragileBoundaries}. Variations of MWPM that deal with hyperedges on the surface code include X/Z correlated decoding \cite{Delfosse2014CorrDecoding} and hybrids of belief-propagation (BP) with MWPM \cite{Fowler2013OptCorrErrs,Paler2023pipelinedcorrelated,Rowshan2026BiasAwareBP,Higgott2023FragileBoundaries}. The X/Z correlated decoder demonstrates improved performance under depolarizing noise on $CSS$ (Calderbank-Shor-Steane) codes \cite{calderbank1996good, steane1996multiple}, but, it is possible to improve this method under biased noise by taking advantage of the asymmetric noise distribution. This has been done for other kinds of decoders, creating bias-aware belief propagation (BP) decoders which capitalize on noise asymmetry, and are effective under code-capacity level noise \cite{Rowshan2026BiasAwareBP}. In this work, we aim to improve the decoding performance on Clifford deformed compass codes by using a correlated MWPM decoder that is able to take the biased noise model into account. 
  
We assess different correlated decoding techniques under biased noise models on $CSS$ and Clifford-deformed elongated compass codes.
In particular, we use a two-stage variant of MWPM that uses joint information caused by one error mechanism to update the matching graph between rounds. We consider two classes of correlated decoding: (i) CSS-only two pass MWPM at code-capacity \cite{Delfosse2014CorrDecoding}, and (ii) hyperedge-informed MWPM at the circuit-level \cite{higgott2022pymatching}. The first takes into account the effects of $Y$ errors on the elongated compass codes under code-capacity noise with additional considerations for noise bias. We call this decoding technique $CSS$ correlated MWPM. For circuit-level analysis, we utilize \texttt{PyMatching}'s correlated decoder \cite{higgott2022pymatching,Higgott2023FragileBoundaries}. We observe that Clifford deformations and compass code stabilizer asymmetry amplify the utility of correlated decoding. Both the $CSS$ and Clifford-deformed compass codes produce structured correlations that exploit two-pass MWPM at a realistic range of noise biases.

This paper is structured as follows. We introduce the elongated compass codes and describe the syndrome extraction circuit we use for circuit-level simulations in Section \ref{sec:codes_noise_model}. Furthermore, we describe the noise models we assume in Section \ref{subseq:noise}. In Section \ref{sec:decoding}, we provide a general description of the MWPM decoder and introduce the correlated decoders we implement. We present the results on code-capacity and circuit-level simulations using correlated decoders in Section \ref{sec:results}.

\section{Codes and Noise Model} \label{sec:codes_noise_model}
\subsection{Elongated Compass Codes}\label{subseq:elong_CC}
Compass codes are subsystem stabilizer codes whose complete gauge group is generated by the interaction operators of the Hamiltonian quantum compass model defined on the square lattice \cite{dorier2005quantum}. The full gauge group is $\{X_{i,j}X_{i,j+1},Z_{i,j}Z_{i+1,j}\}$ where $i (j)$ is the row (column) index of the qubit. Starting from this gauge group, we can create stabilizer codes by fixing different sets of gauge operators \cite{li20192d,huang2020fault,bombin2015gauge,paetznick2013universal}. 


Elongated compass codes are a family of codes whose stabilizers are obtained by fixing a set of gauge operators of the compass codes according to the elongation parameter $\ell$ \cite{li20192d}. Suppose we label the cells of the $d\times d$ square lattice with coordinates $1\leq i,j \leq n -1$. Then we fix the $X$ gauge operators surrounding any cell whose coordinates satisfy $i - j \equiv 0 \mod \ell$. The remaining $Z$ gauge operators between the weight-4 $X$ stabilizers are fixed to produce $Z$ stabilizers of weight $2\ell$ in each row. To ensure commutativity, we fix the remaining weight-2 $X$ gauge operators surrounding the $Z$ stabilizers. Note that when $\ell=2$, we create the rotated surface code \cite{bombin2007optimal}. The distance of the resulting code is $d$, the dimension of the lattice, indicating that we can correct up to $\lfloor \frac{d-1}{2}\rfloor$ errors. 

Note that elongated compass codes are $CSS$ codes. Their stabilizer generators can be divided into a set of stabilizers which are products of only Pauli-$X$ and another set which are a product of only Pauli-$Z$ \cite{calderbank1996good,steane1996multiple}. We refer to each of these sets as $X$ stabilizer and $Z$ stabilizers, respectively. These codes are useful when considering biased dephasing errors. As $\ell$ increases, the number of weight-2 $X$ stabilizers increases, gathering more accurate information on the location of dominant dephasing errors. This property comes at the cost of reduced $X$-error sensitivity, leading to a trade-off in the performance of the $X$ and $Z$ error correction. This trade-off is balanced by some optimal bias at which the total threshold is maximized. The optimal biases and maximum thresholds increase with elongation $\ell$, indicating that codes with higher elongation perform better under noise models with higher biases. 

\subsection{Clifford Deformations}\label{subseq:CD}
One technique used to improve the performance of stabilizer codes are Clifford deformations \cite{bonilla2021xzzx,dua2022clifford,debroy2021optimizing,campos2025}. The Clifford deformation of a stabilizers code is the application of a set of Clifford transformations on the code space which modifies the basis of the stabilizers without changing the support or weight of the original stabilizers \cite{dua2022clifford}. A popular example is the $XZZX$ surface code which applies Hadamard transformations on every other qubit of the surface code lattice \cite{bonilla2021xzzx}. This code achieves higher thresholds than the $CSS$ surface code under biased Pauli noise due to the symmetries imposed by the Clifford deformation \cite{bonilla2021xzzx,tuckett2019tailoring, Tuckett2020FTSCBias}. 

\begin{figure}
    \centering
    \includegraphics[width=0.45\linewidth]{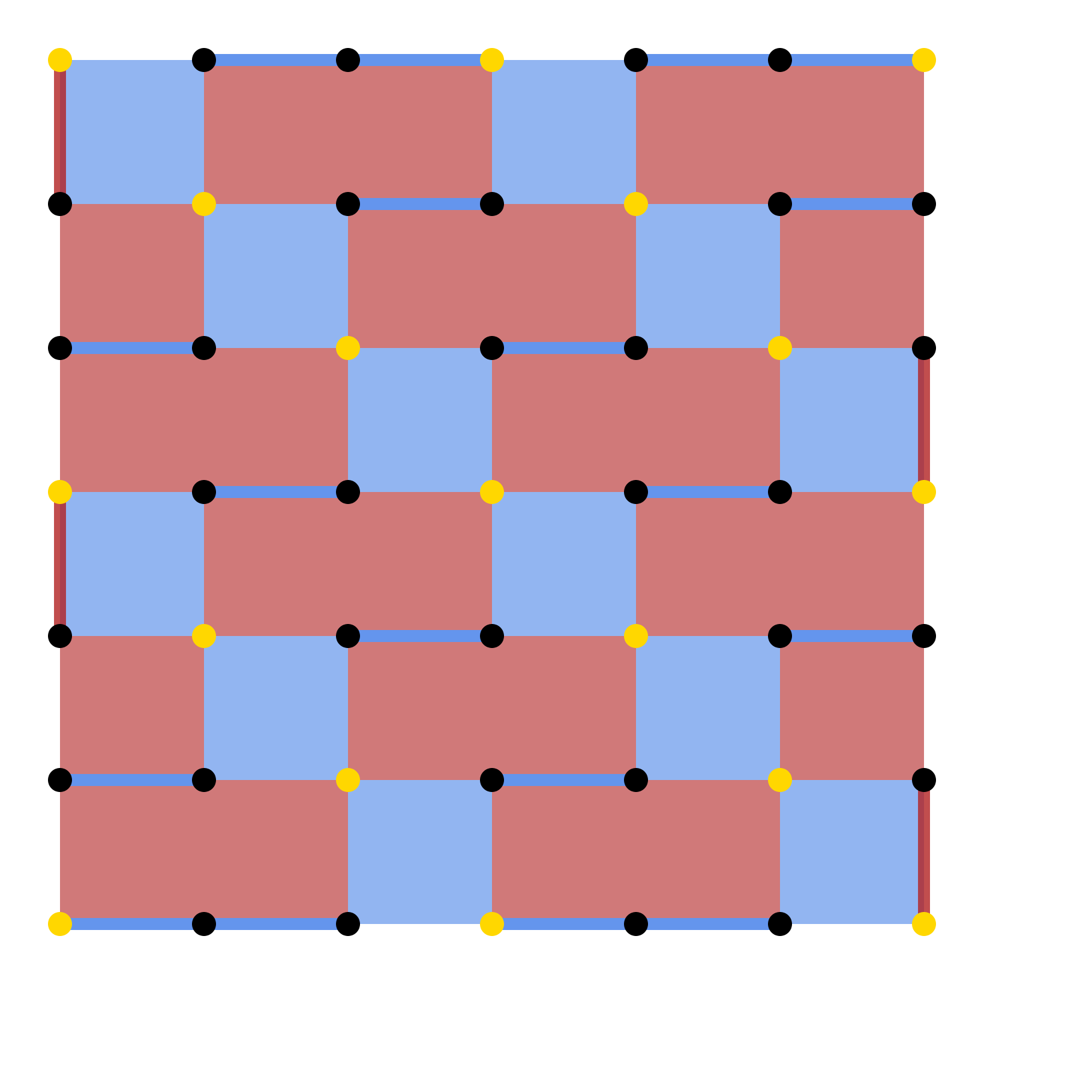}
    \caption{Schematic of an elongated compass code with elongation $\ell=3$ of distance $d=7$. $X$ stabilizers are blue, $Z$ stabilizers are red. The qubits are at the vertices of the lattice. We apply the $ZXXZ\square$ Clifford deformation on this code by applying Hadamard transformation on the yellow qubits. The weight-2 $X$ stabilizers detect the location of $Z$ errors with high accuracy, leading to larger thresholds under noise models with dominant $Z$ errors.}
    \label{fig:cd_elong_cc}
\end{figure}

In \cite{campos2025}, the authors found Clifford deformations that improved the performance of the elongated compass codes introduced in Section \ref{subseq:elong_CC}. One of the deformations explored was the $ZXXZ \square$ deformation, which applies Hadamard transformations on the top left and bottom right qubits of the weight-4 $X$ stabilizers of the elongated compass codes. Note that for $\ell = 2$, $ZXXZ\square$ deformation is equivalent to the $XZZX$ surface code. In Figure \ref{fig:cd_elong_cc}, we show the $ZXXZ\square$-deformed elongated compass code with elongation $\ell = 3$. The resulting $ZXXZ\square$-deformed codes demonstrated the best performance of the codes considered. In this work, we extend the analysis of the $ZXXZ\square$-deformed elongated compass codes to consider circuit-level noise. 

\subsection{Syndrome Extraction Circuit}\label{subseq:syndrome_ext_circuit}
We perform code-capacity and circuit-level simulations on $CSS$ and $ZXXZ\square$-deformed elongated compass codes. To complete $X(Z)$-type memory experiments, we measure $X(Z)$ stabilizers to check for errors. Each memory experiment consists of $d$ rounds of repeated syndrome extraction used to calculate the logical error rates for $X$ and $Z$ memories under circuit-level noise. A logical error is counted when an $X$ or $Z$ logical error occurs, or when both occur.  

\begin{figure}[!t]
    \centering
    \subfloat[]{\includegraphics[width=0.7\linewidth]{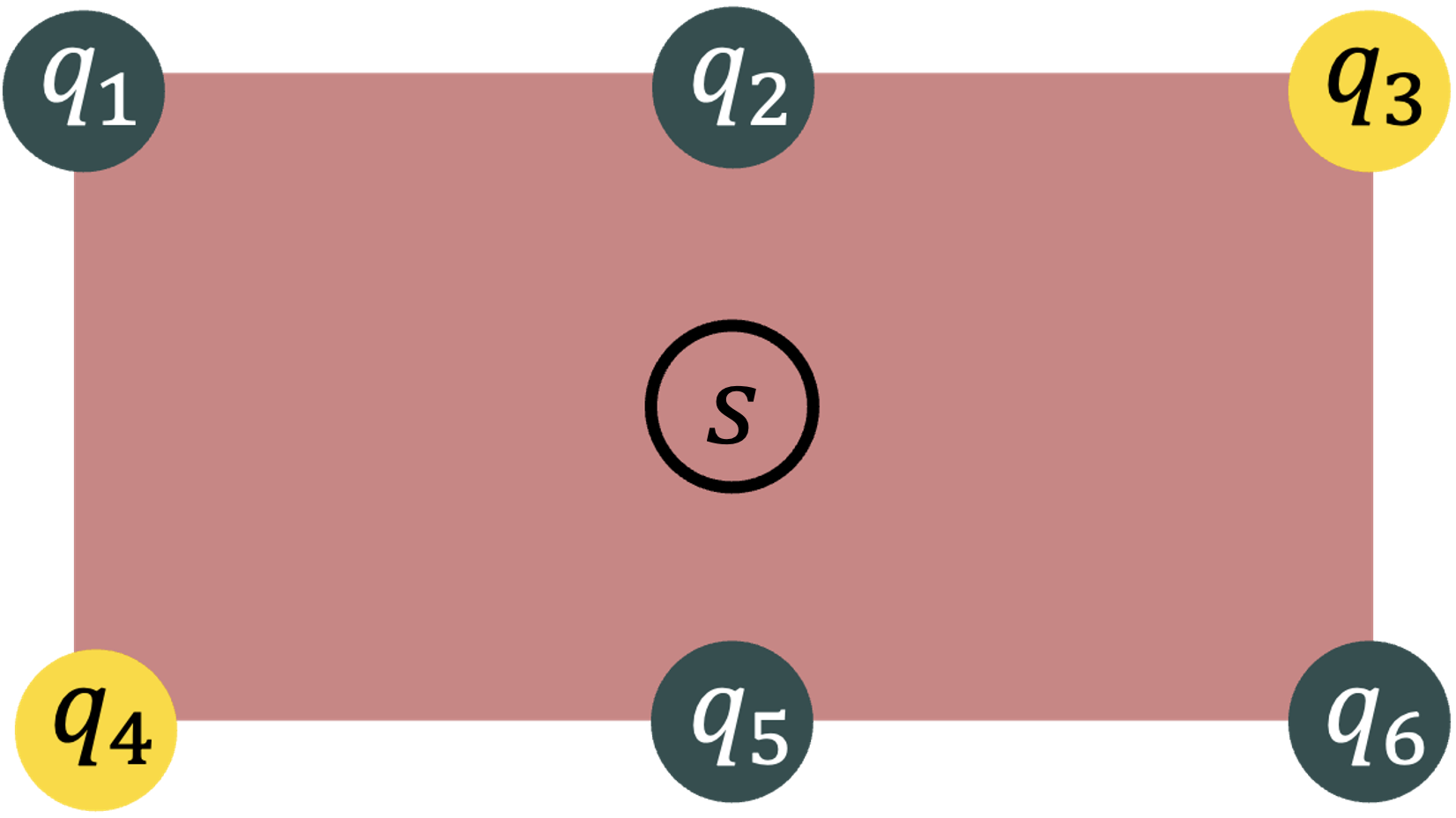}\label{fig:stab_drawing_forcircuit}}
    \\
    \subfloat[]{\begin{quantikz}[column sep=1.3em, row sep=1.1em]
        \lstick{$s$} & \gate{H} & \ctrl{1}  & \ctrl{4} & \ctrl{2}  & \ctrl{5} & \ctrl{3} & \ctrl{6} & \gate{H} & \gate{M}\\
        \lstick{$q_1$} & \qw      & \ctrl{-1} & \qw      & \qw       & \qw      & \qw      & \qw      & \qw & \qw \\
        \lstick{$q_2$} & \qw      & \qw       & \qw      & \ctrl{-2} & \qw      & \qw      & \qw      & \qw & \qw \\
        \lstick{$q_3$} & \qw      & \qw       & \qw      & \qw       & \qw      & \targ{}  & \qw      & \qw & \qw \\
        \lstick{$q_4$} & \qw      & \qw       & \targ{}  & \qw       & \qw      & \qw      & \qw      & \qw & \qw \\
        \lstick{$q_5$} & \qw      & \qw       & \qw      & \qw       & \ctrl{-5}& \qw      & \qw      & \qw & \qw \\
        \lstick{$q_6$} & \qw      & \qw       & \qw      & \qw       & \qw      & \qw      & \ctrl{-6}& \qw & \qw 
        \end{quantikz}
        \label{fig:synd_ext_circuit}}
    \caption{\ref{fig:stab_drawing_forcircuit}) Depiction of a weight-6 $Z$ stabilizer of $ZXXZ\square$-deformed elongated compass code with an elongation of $\ell=3$. Data qubits ($q_i$) and syndrome qubit ($s$) are labeled. Data qubits are colored yellow if they undergo a Hadamard transformation. \ref{fig:synd_ext_circuit}) Circuit diagram of syndrome extraction for stabilizer drawn in \ref{fig:stab_drawing_forcircuit}. We do not include the readout stage in this figure. Readout in our circuit happens for each syndrome qubit at the end of each of $d$ rounds.}
\end{figure}

We use \texttt{Stim} \cite{Gidney2021stimfaststabilizer} to perform QEC circuits and sample errors for circuit-level simulations. Under circuit-level noise, a QEC cycle includes $d$ rounds of repeated syndrome extraction, where $d$ is the distance of the code being tested. The steps for each QEC cycle are as follows: (i) Initialize the logical state to the $X(Z)$ basis when performing $X(Z)$ memory experiment. (ii) Perform syndrome extraction serially for each stabilizer (see Fig. \ref{fig:synd_ext_circuit}). (iii) Measure the syndrome qubits. (iv) Repeat steps (ii) and (iii) $d$ times. (v) Measure all qubits and extract the $X(Z)$ logical of interest.

\subsection{Noise Model}\label{subseq:noise}
In code-capacity simulations, we use an asymmetric Pauli channel to define our noise model as shown in Eq. \ref{eq: asym_pauli_channel} \cite{tuckett2018ultrahigh,tuckett2019tailoring,Martinez2026leveragingbiasatcircuitlevel}. The probability of error on one physical qubit is $p = p_x + p_y + p_z$ where $p_x, p_y$ and $p_z$ correspond to the probabilities of Pauli error $X, Y$ and $Z$ on a qubit, respectively. 

\begin{equation}
    \label{eq: asym_pauli_channel}
    \mathcal{E}[\rho] = (1-p)\rho \ + \ p_x X\rho X \ + \ p_y Y\rho Y \ + \ p_z Z \rho Z
\end{equation}

We quantify the bias towards dephasing errors with the parameter $\eta = \frac{p_z}{p_x + p_y}$ and assume that $p_x = p_y$. A noise bias of $\eta = 0.5$ represents equal $X$ and $Z$ noise \cite{tuckett2019tailoring, Martinez2026leveragingbiasatcircuitlevel}.


To simulate biased circuit-level noise, we implement a version of the hybrid biased-depolarizing (HBD) noise model introduced in \cite{Martinez2026leveragingbiasatcircuitlevel} with a modification on idling errors and state preparation and measurement (SPAM) errors. For bias-preserving $CZ$ gates, we apply a two-qubit biased Pauli channel with probability $p$ and bias $\eta$. In this case, pure dephasing errors ($IZ,ZI,ZZ$) occur with probability $\frac{\eta p}{3(1+\eta)}$ and all other errors with probability $\frac{p}{12(1+\eta)}$. $CNOT$ and $H$ gates do not preserve bias, so we apply two- and single-qubit depolarizing channels after each gate \cite{Martinez2026leveragingbiasatcircuitlevel}. We apply idling noise on all qubits between rounds of memory with single-qubit asymmetric Pauli channel $p_x = \frac{p}{2(1+\eta)}$ and $p_z = \frac{\eta p}{1 +\eta}$ as defined for code-capacity noise above. Applying idling errors between rounds allows us to simplify circuit design and treat the measurement serially \cite{li20192d}. 
Measurement errors occur with probability $p$. 

\section{Decoder Construction}\label{sec:decoding}
\begin{figure*}
    \centering
    \includegraphics[width=0.75\linewidth]{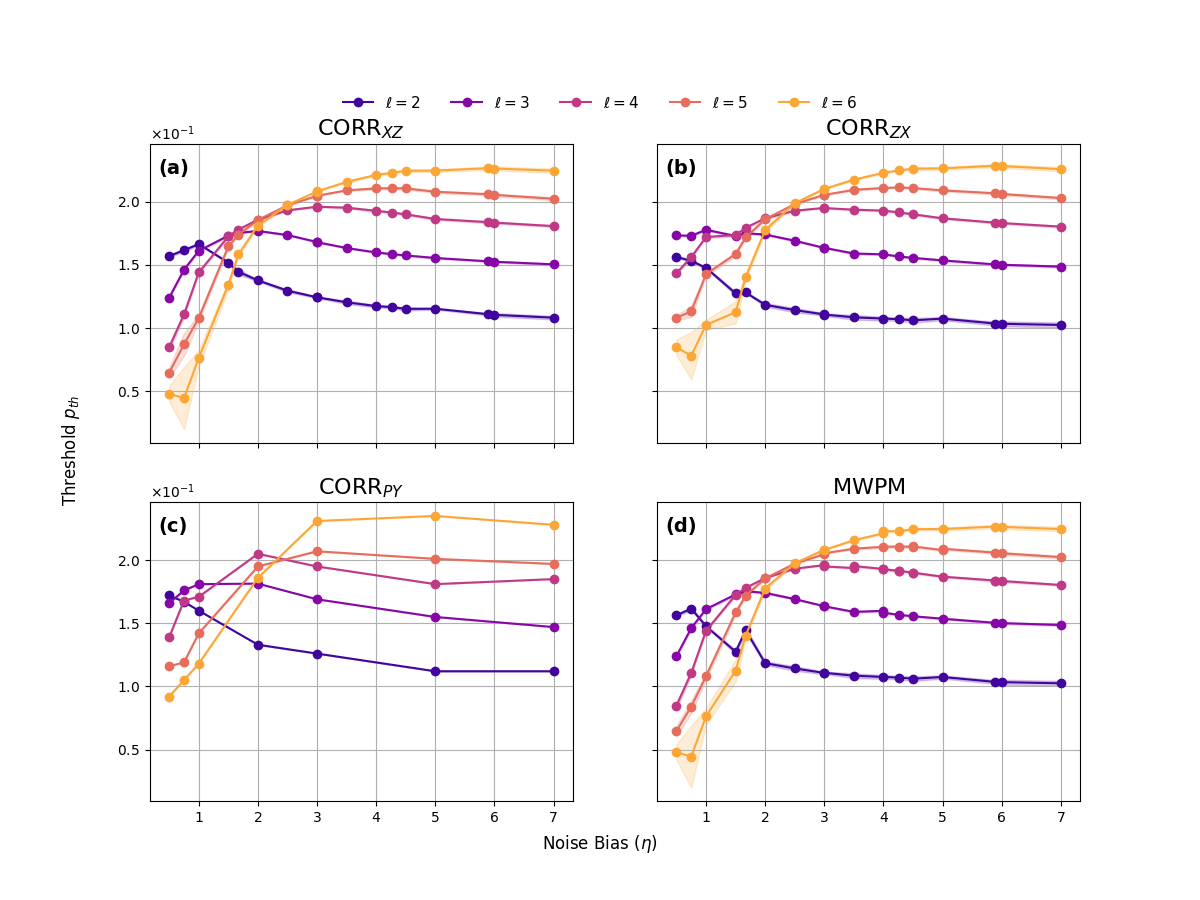}
    \caption{$CSS$ correlated decoding in (a), (b) compared to \texttt{PyMatching} correlated decoding in (c) and MWPM in (d). (a) The thresholds of the compass codes with $CSS$ correlated decoding over bias $\eta$ correcting $X$ errors, then $Z$ errors. (b) The thresholds for $CORR_{ZX}$ ($Z$ error then $X$ error) decoding at each $\ell$. (c) \texttt{PyMatching} correlated decoder thresholds over a range of $\eta$. To achieve a similar to code-capacity decoding scheme with the \texttt{PyMatching} correlated decoder, we use $d$ rounds of memory with the Pauli channel in section \ref{sec:codes_noise_model} acting on each data qubit before the rounds (d) Standard MWPM thresholds compared to the bias $\eta$ for different $\ell$.  
    }
    \label{fig:code_cap_vector_vs_hyperedge}
\end{figure*}
Our decoding analysis is based on the minimum weight perfect matching (MWPM) algorithm \cite{edmonds1965paths}. The MWPM algorithm takes in a matching graph $\mathcal{G} = (\mathcal{V}, \mathcal{E}, \mathcal{W})$ where $\mathcal{V} = \{v_i\}$ are the vertices representing detectors in our circuit, and  $\mathcal{E} = \{(v_i, v_j)\}$ are the edges representing qubits in our lattice with error rates given by $\mathcal{W} = \{w_{ij}\}$ between detectors $v_i$ and $v_j$. For a probability of error $p$ on a given qubit, the weight is given by $w_{ij} = \log(\frac{1-p}{p})$ \cite{higgott2022pymatching}. The output of the algorithm is the minimum weight error path that would turn on the provided detection events.  

\subsection{Correlated Decoding}\label{subseq:corr_MWPM}
\subsubsection{$CSS$ Correlated Decoding}
We implement a version of X/Z correlated MWPM that is tailored to a biased noise model \cite{Delfosse2014CorrDecoding}. The conventional approach to decoding $CSS$ codes is to assume that $X$ and $Z$ errors occur independently, neglecting the correlations between these due to $Y = XZ$, up to an overall phase. We complete the $CSS$ correlated decoding process by separately decoding $X$ and $Z$ errors using MWPM. We incorporate the correlations that arise due to $Y$ errors by setting edge probabilities to the conditional probabilities for an error $\mathcal{E}=\mathcal{E}_1\mathcal{E}_2$:
\begin{equation}
\label{eq:cond_probs}
\begin{array}{c}
     P(\mathcal{E}_2 = X | \mathcal{E}_1 = Z) = \frac{1}{1 + 2\eta}  \\
     P(\mathcal{E}_2 = Z | \mathcal{E}_1 = X) = \frac{1}{2}
\end{array}
\end{equation}
These conditional probabilities can be passed between $X$ and $Z$ decoding steps. Since we consider an asymmetric noise channel, we explore how the ordering of this decoding method affects decoding accuracy. We consider $XZ$ ordering where we decode $X$ errors ($Z$ syndromes) first and use the results to inform the decoding of $Z$ errors ($X$ syndromes). Similarly, we consider the reversed $ZX$ ordering.

In the $XZ$ correlated decoding process, the first step is to decode $X(Z)$ errors using MWPM. Given the correction on the $X(Z)$ errors, we update the matching graph for $Z(X)$ errors. We set the weight of the edges in correction $w_e = \log{\frac{1-p_c}{p_c}}$ in the $Z(X)$ matching graph for the second stage of decoding. $p_c$ is the conditional probability of an error on that edge given an $X(Z)$ error occurred in the first round of decoding. For $X$ then $Z$ error decoding ($XZ$ decoding), use $p_c = P(\mathcal{E}_2 = X | \mathcal{E}_1 = Z)$. If the decoding order is $ZX$, then we use $p_c = P(\mathcal{E}_2 = Z | \mathcal{E}_1 = X)$ from equation \ref{eq:cond_probs}. To complete the next decoding step, we use the conditional weight-adjusted MWPM graph to decode $Z(X)$ errors. 


\subsubsection{Circuit-level Correlated Decoding}
We implement the correlated MWPM decoder with \texttt{PyMatching} \cite{higgott2022pymatching}. The matching graphs are created from the detector error model (DEM) generated by \texttt{Stim} corresponding to our QEC circuits \cite{ Gidney2021stimfaststabilizer}. DEMs record all possible error mechanisms encoded by the circuit, while tracking parity of measurements included in detectors. Similar to matching graphs, a DEM can be represented by a weighted graph (or hypergraph) whose edges correspond to error mechanisms and nodes represent detectors. The hyperedges from the DEM must be decomposed into edges to create a matching graph on which we can perform MWPM. Edge weights are given by the log-likelihood ratios derived from the error probabilities in the DEM.

At the circuit-level, not all correlations are caused by $Y$ errors. To consider circuit-level correlations, we use \texttt{PyMatching}'s correlated decoder \cite{higgott2022pymatching}. We use a similar strategy to code-capacity to capture circuit-level correlations. When decomposing hyperedges to go from a DEM hypergraph representation to a matching graph, some information is lost \cite{Higgott2023FragileBoundaries}. This information is utilized by completing two passes of MWPM.

By associating each error mechanism with the corresponding decomposed hyperedges, we can use an initial pass of MWPM to update the weights of the matching graph. The strategy of associating correlations captured by hyperedges is correlated decoding. Each hyperedge consists of edges, with joint error probability $P(E_1\cap E_2\cap ...)$, where each edge has marginal error probability $P(E_1), P(E_2), ...$. Now consider a hyperedge decomposable into two edges, $E_1$ and $E_2$. For an edge $E_1$ included in the correction, the conditional probability $p_c$ analogous to the code-capacity case is $p_c = P(E_2 | E_1) = \frac{P(E_1 \cap E_2)}{P(E_1)}$. This conditional probability is then used to adjust the weight of $E_2$ in the matching graph. 

\section{Results and Analysis}\label{sec:results}
\begin{figure*}
    \centering
    \includegraphics[width=\linewidth]{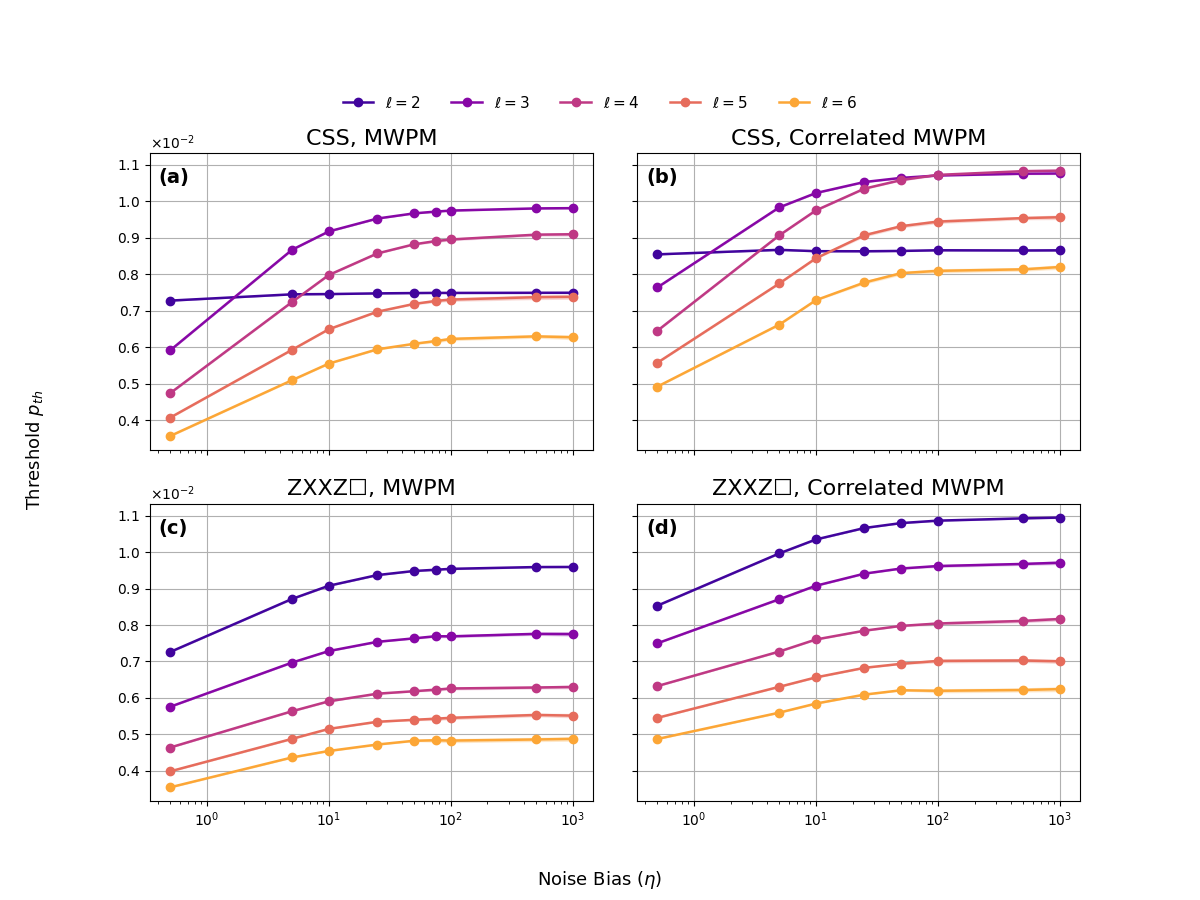}
    \caption{The circuit-level correlated decoding of the Clifford-deformed elongated compass codes and the standard MWPM directly compared over $\eta$. (a) The $CSS$ elongated compass codes threshold compared to bias $\eta$ with MWPM, and (b) with correlated MWPM. (c), (d) The thresholds of the $ZXXZ\square$-deformed code over a range of $\eta$ with MWPM in (c) and correlated MWPM in (d). The surface code ($\ell = 2 \ CSS$) experiences no change with $\eta$.}
    \label{fig:pycorr_dec_thr_vs_eta}
\end{figure*}
\subsection{Circuit-level Clifford Deformed Elongated Compass Code Thresholds MWPM}\label{subseq:CD_cc_thresholds}
We analyze the performance of correlated decoding on $CSS$ and $ZXXZ\square$-deformed elongated compass codes by determining their thresholds under circuit-level noise with varying bias values ($\eta$). The noise model implemented is described in Section \ref{subseq:noise}. We determine the thresholds by running simulations of each code with distances $d=11,13,15,17,19$ to record logical error rates given some input physical error rate. From the resulting threshold plots, we assume that the curves are quadratic in $x = (p-p_{th})d^{1/\nu}$ near the threshold. The parameters here correspond to the threshold ($p_{th}$), the distance ($d$), and the critical exponent ($\nu$) \cite{wang2003confinement}. We extract threshold values from quadratic fits for codes with elongations $\ell \in\{ 2,3,4,5,6\}$ under noise models with biases of $\eta \in \{0.5, 5, 10, 25,50,100,500,1000\}$. All threshold values and data to reproduce results is accessible in \cite{correlated_decoding_cd_compass_codes_2026, meinking_code_2026}.

In Figure \ref{fig:pycorr_dec_thr_vs_eta}, we plot the thresholds of the $CSS$ and $ZXXZ\square$-deformed codes when using standard MWPM and the correlated MWPM. For both decoding methods, we observe that the deformation improves thresholds of the $\ell = 2$ elongated compass code as the bias increases. However, with increasing $\ell$, the advantage of the $ZXXZ\square$ deformation over the $CSS$ codes is lost. In addition to the high stabilizer weight, this is most likely due to the fact that all stabilizers include non bias-preserving gates. The $Z$-stabilizer syndrome circuits of the $CSS$ codes are all bias preserving.

The syndrome extraction circuits of the $ZXXZ\square$-deformed elongated compass code stabilizers involve both CNOT and CZ gates. The mixture of bias-preserving (CZ) and depolarizing (CNOT, H) reduces the assumed bias $\eta$ at circuit-level and produces a plateau in the thresholds of the $ZXXZ\square$ codes instead of the increasing thresholds observed at the code-capacity noise level \cite{campos2025}. The thresholds of the $CSS$ codes retain threshold improvements from code-capacity simulation. We continue using the gate error framework of the HBD noise model to retain experimental applicability, despite the bias saturation effect. To counteract this effect and improve thresholds at all $\eta$, we apply correlated decoding. 

\subsection{Decoding Improvements in Threshold}\label{subseq:corr_thresholds} 
\begin{figure*}
    \centering
    \includegraphics[width=\textwidth]{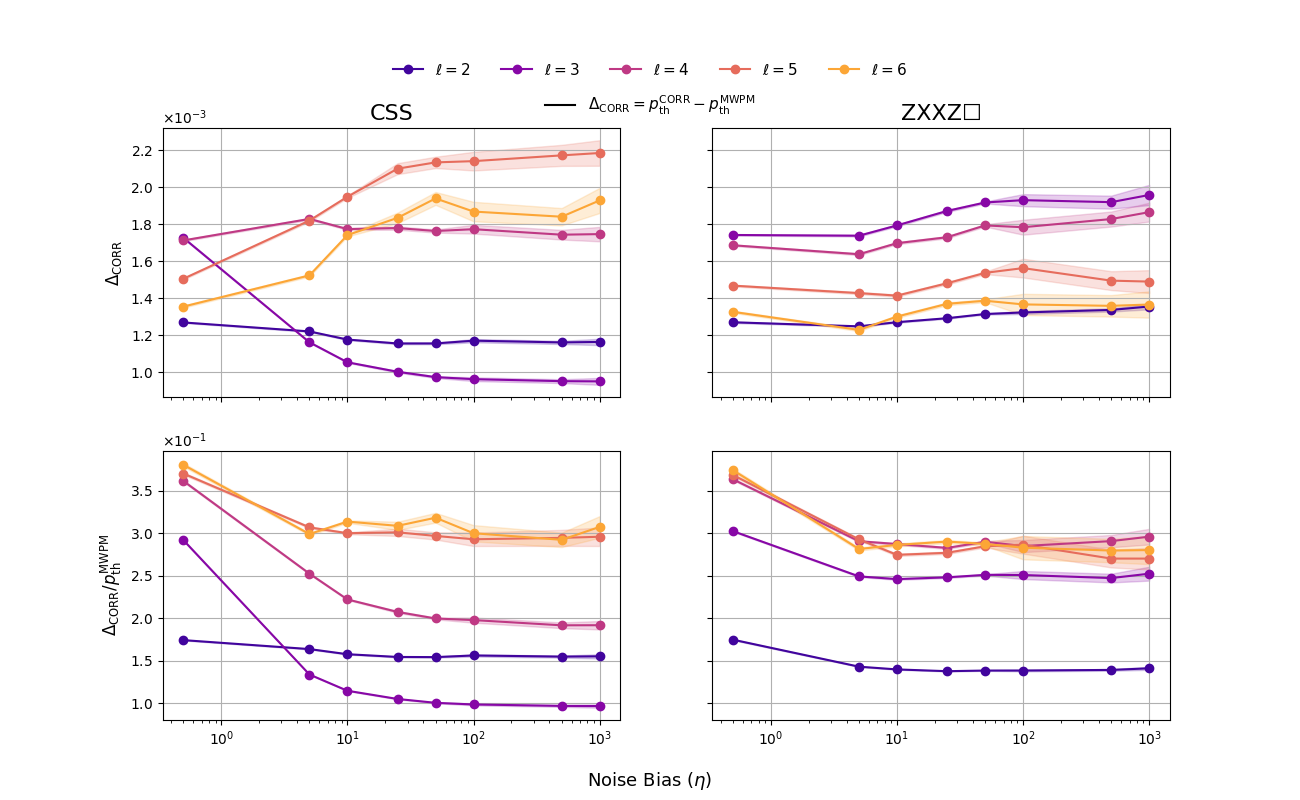}
    \caption{Threshold gain ($\Delta_{CORR}$) and relative gain ($\Delta_{CORR}/p^{MWPM}_{th}$) of correlated decoding compared to standard MWPM over a range of biases $\eta$ for $CSS$ and $ZXXZ\square$-deformed elongated compass codes. The $CSS$ codes experience tradeoffs in threshold since high stabilizer asymmetry is preferred to maximize correlated decoding gains. The $\Delta_{CORR}$ of  $ZXXZ\square$ codes with increasing $\eta$, though the maximum gain is achieved by the $CSS$ codes. The relative correlated decoding gain, $\Delta_{CORR}/p^{MWPM}_{th}$, plateaus at high $\eta$ for both the $CSS$ and $ZXXZ\square$-deformed codes. The relative gain of the correlated decoder on the $ZXXZ\square$-deformed compass codes improves with elongation for all biases.}
    \label{fig:deltacorr_vs_bias}
\end{figure*}
At code-capacity, the benefits of correlated decoding are suppressed as the bias increases, as shown in Figure \ref{fig:code_cap_vector_vs_hyperedge}. The advantage of code-capacity correlated decoding is that it more effectively decodes the correlations due to $Y$ errors. As we increase $\eta$, the relative frequency of $Y$ errors decreases, reducing the advantage of the correlated decoder. In particular, we observe that the thresholds achieved by the correlated decoder converge to those of the standard MWPM when $\eta \gtrsim 2$ in Figure \ref{fig:code_cap_vector_vs_hyperedge}. 

Additionally, we note that the decoding order affects performance. To choose the optimal order and properly take advantage of code-capacity correlated decoding, we must keep in mind the information imbalance of $Z$ and $X$ matching graphs due to stabilizer and noise asymmetry \cite{Delfosse2014CorrDecoding}. Since the asymmetry favors the detection of $Z$ errors by the $X$ stabilizers, it is expected that decoding $Z$ errors, then $X$ errors will utilize correlated decoding best. This is supported by our results shown in Figure \ref{fig:code_cap_vector_vs_hyperedge} where subfigure b) demonstrates higher thresholds for biases $\eta\lesssim2$. 

Since we seek performance improvements over a wide range of $\eta$, we directly compare our code-capacity $CSS$ correlated decoder (see Sec. \ref{sec:decoding}), and \texttt{PyMatching}'s correlated decoder on the $CSS$ elongated compass codes. In Figure \ref{fig:code_cap_vector_vs_hyperedge}, we find that \texttt{PyMatching}'s correlated decoder achieves thresholds comparable to those of the $CSS$ correlated decoder with $ZX$ ordering. These two methods take advantage of the noise model. The $ZX$ ordered $CSS$ correlated decoder follows the information imbalance of the stabilizers of the code, and accounts for $\eta$ when updating edge weights with $p_c$. The \texttt{PyMatching} correlated decoder is directly informed by the error mechanism probabilities contributing to the hyperedges of the DEM, enabling more accurate decoding. This advantage of the \texttt{PyMatching} correlated decoder is translated to biased circuit-level noise. 

At the circuit-level, we observe that correlated decoding increases the thresholds of both $CSS$ and $ZXXZ\square$-deformed elongated compass codes for all biases considered compared to standard MWPM, as shown in Figure \ref{fig:pycorr_dec_thr_vs_eta}. Note that these improvements are much more significant compared to those achieved at the level of code-capacity noise, particularly at high bias values. This is a consequence of the fact that circuit simulations exhibit many more correlations than code capacity simulations, which the \texttt{PyMatching} correlated decoder can take into account. To quantify the impact of correlated decoding in comparison to standard MWPM, we define $\Delta_{CORR} = p^{CORR}_{th} - p^{MWPM}_{th}$ where $p^{CORR}_{th}$ is the threshold achieved by the \texttt{PyMatching} correlated decoder and $p^{MWPM}_{th}$ is that achieved by the standard MWPM decoder. Furthermore, we define the relative improvement of correlated decoding, $\Delta_{CORR}/p^{MWPM}_{th}$.

We plot these quantities with respect to bias in Figure \ref{fig:deltacorr_vs_bias}. Generally, we observe that the improvement of the correlated decoder on $CSS$ elongated compass codes appears to have asymptotic behavior as bias increases. In contrast, the absolute gain in thresholds of the correlated decoder on the $ZXXZ\square$-deformed codes improves with bias. Note, however, that the improvements are greater for $CSS$ codes with higher $\ell$ whereas the opposite is true for the $ZXXZ\square$-deformed codes with the exception of the rotated surface code ($\ell = 2$). The relative gain in the thresholds of codes with low elongation values could be limited for due to insufficient asymmetry that could be exploited during the decoding of biased errors.

For $CSS$ codes, $\Delta_{CORR}$ increases for $\ell > 4$ with bias. The maximum relative gain $\Delta_{CORR}/p^{MWPM}_{thr} = 35\%$ is achieved by the elongated compass code with $\ell=6$ with $\eta= 0.5$. However, the $\Delta_{CORR}$ for the $CSS$ elongated compass code of $\ell=3$ decreases with $\eta$. We hypothesize the highest relative gains are achieved by codes with high elongation due to the bias preservation of the syndrome extraction circuits of $Z$ stabilizers. 


Unlike the the $CSS$ case, the $ZXXZ\square$ codes have relative gain $\Delta_{CORR}/p^{MWPM}_{th}$ increasing with elongation for all $\eta$. We expected the $ZXXZ\square$ deformed codes to produce the largest $\Delta_{CORR}$ due to symmetric regions in the code-capacity matching graphs from previous work \cite{campos2025}. However, depolarizing gate errors break these symmetric regions under our circuit-level noise model. This suggests $\Delta_{CORR}$ may increase with a bias-preserving noise model; verifying this remains future work.  

To test decoder confidence with increasing $\eta$, we compute complementary gaps for the \texttt{PyMatching} correlated decoder and standard MWPM. See appendix \ref{sec:appendixc_comp_gap}.

\section{Conclusion}\label{sec:conclusion}

We have demonstrated that correlated decoding improves the performance of Clifford-deformed compass codes under biased circuit-level noise. Correlated MWPM outperforms standard MWPM across all bias regimes considered under circuit-level noise. This improvement is highest for codes with a large elongation parameter, highlighting the decoder's ability to leverage information on correlated errors from high-weight stabilizers. Bias-tailoring introduces an advantage in $CSS$ two-stage correlated decoding at code-capacity. Our choice of noise model causes the thresholds of all codes we consider to saturate at high $\eta$ due to the lack of bias preservation. Notably, this removes the advantage of the $ZXXZ\square$ for $\ell > 2$ over the $XZZX$ surface code that had been observed in code-capacity noise \cite{campos2025}. 

Correlated decoding may offer more benefits for these codes than threshold increases. Fragile boundaries \cite{Higgott2023FragileBoundaries} are a problem for Clifford deformed codes. The error decomposition done in the correlated decoding process may be a promising method to circumvent distance reductions due to fragile boundaries of Clifford deformed codes under biased noise. 

\section{Acknowledgments}
The authors thank E. Takou and Y. Lin for their insights, debugging help, and thoughtful discussions. This work
was supported by the NSF QLCI for Robust Quantum
Simulation (OMA-2120757) and the Office of the Director of National Intelligence (ODNI), Intelligence Advanced Research Projects Activity (IARPA), under the
Entangled Logical Qubits program through Cooperative
Agreement Number W911NF-23-2-0216.

\appendices
\section{Syndrome Extraction Circuit Gate Order}\label{sec:appendixb_synd_ext_circ}
\begin{figure*}[t!]
    \centering
    \includegraphics[width=0.8\linewidth]{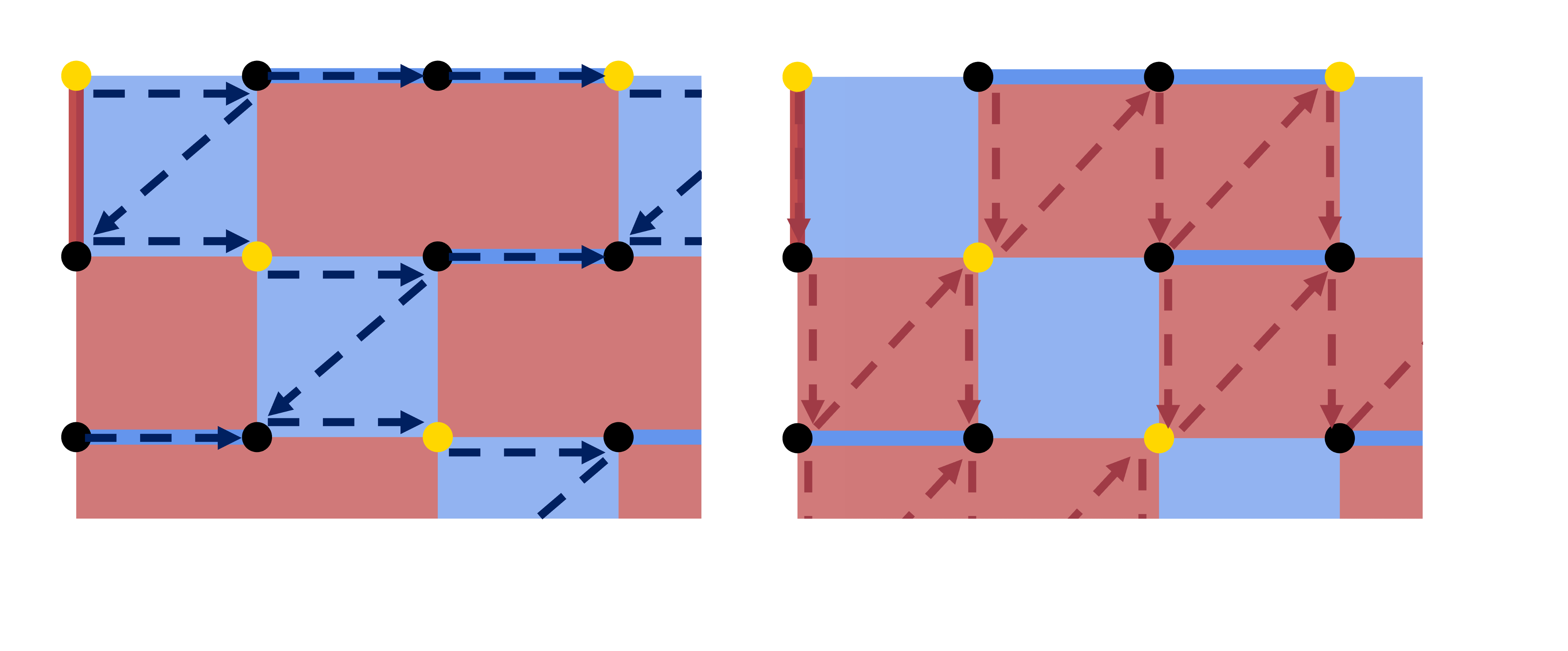}
    \caption{The order of the CNOT/CZ gates in the extraction step of the circuit. The order is picked as to avoid propagating hook errors and ensure stabilizers commute. The ordering is chosen independent of applied Hadamard transforms. (Left) The $X$-memory experiment measurement order. The blue are the $X$-stabilizers, and in the serial form of the circuit we measure each column of the code from top to bottom, proceeding left to right. (Right) The $Z$-memory experiment measurement order. The red are the $Z$-stabilizers. We measure each row of the code from left to right, iterating rows from top to bottom.}
    \label{fig:synd_ext_ordering}
\end{figure*}
The sequence of CNOTs/CZs is shown in figure \ref{fig:synd_ext_circuit}. To prevent hook errors and ensure that all of our measurements commute, we must decide an ordering for our gates\cite{tomita2014low,huang2020fault}. The ordering of extraction was originally determined in \cite{huang2020fault}, and remains relevant after completing Clifford deformations since each qubit is in at least two stabilizers. The ordering we implement is shown in Figure \ref{fig:synd_ext_ordering}.

\section{Complementary Gap}\label{sec:appendixc_comp_gap}

\begin{figure*}[ht!]
    \centering
    \includegraphics[width=0.75\textwidth]{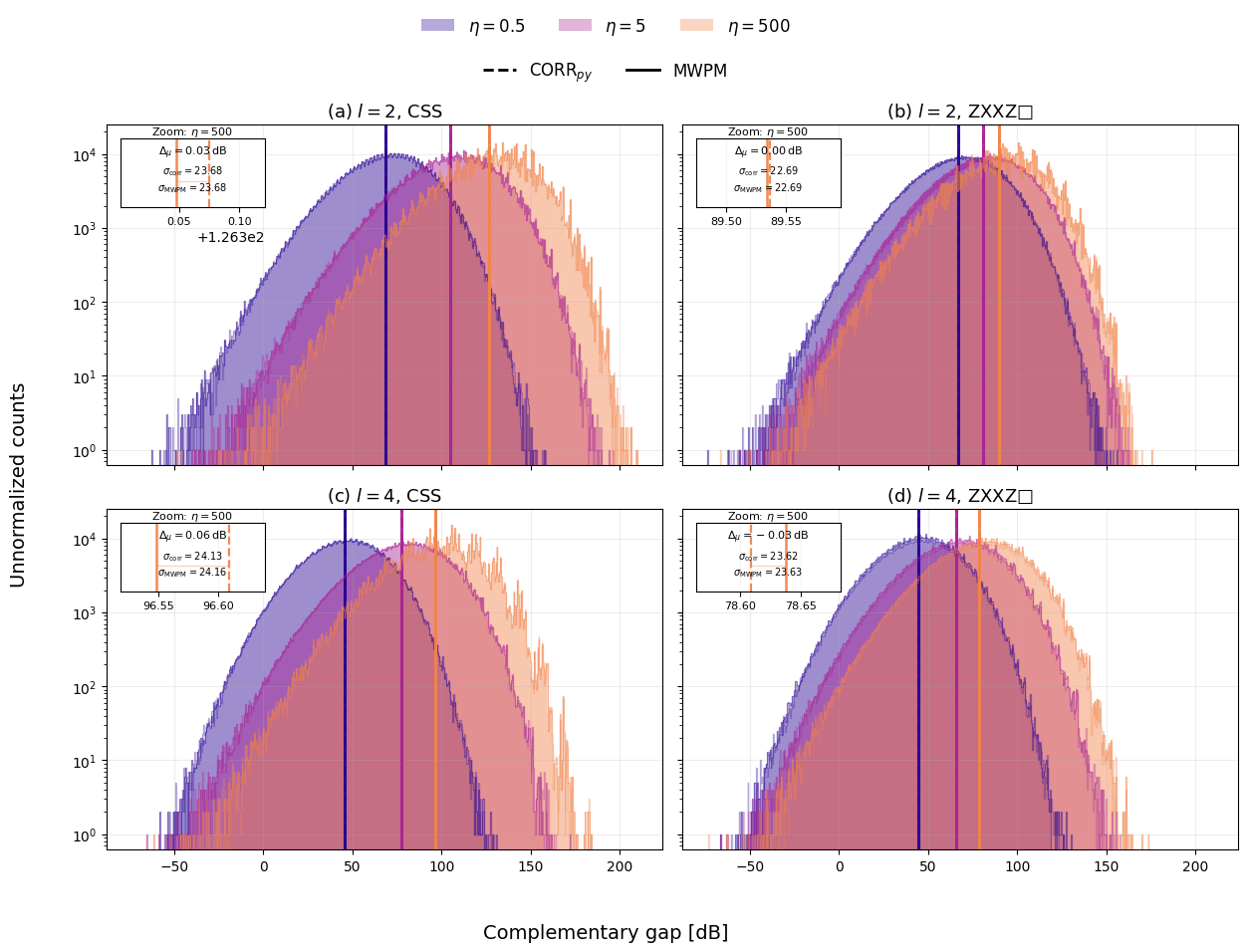}
    \caption{The signed complementary gap histogram distributions for $Z$-memory, $d = 11$ and the noise model described in Section \ref{sec:codes_noise_model} with $p = 0.005$ comparing \texttt{PyMatching} correlated decoding and MWPM. The $x$-axis is scaled in dB as in \cite{Gidney2024YokedSurfaceCodes}, where 1 dB $= -\log{\frac{p}{1-p}}\times\frac{10}{w}$ where $p$ is the probability of error for an edge, and $w$ is its weight. Each color represents a different noise bias, $\eta = \{0.5, 5, 500\}$. The inset shows the mean and variance of one $\eta$ of the complementary gap for \texttt{PyMatching} correlated decoding and MWPM, where the vertical lines represent the means of each distribution. (a),(c) Show the $\ell = 2, \ell=4, CSS$ deformed codes respectively. (b),(d) Show $\ell = 2, \ell=4, ZXXZ\square$ deformed codes. A negative $\Delta$ in this case indicates that the mean of \texttt{PyMatching} correlated decoding was lower than MWPM.}
    \label{fig:gap_over_eta}
\end{figure*}

We assess our decoder confidence at different $\eta$ via the signed complementary gap. The minimum weight error that is picked by a round of MWPM (edges $E_{min}$, weight $W_{min}$) has a complementary error chain (edges $E_{comp}$, weight $W_{comp}$) which combine to a logical error \cite{Gidney2024YokedSurfaceCodes,Toshio2025DecoderSwitching}. The difference in the weights of these two errors is known as the "complementary gap". Explicitly, this gap is $g = W_{comp} - W_{min}$. A negative signed gap indicates the minimum weight correction applied by the decoder incorrectly counted logical parity. Decoder confidence is represented by the magnitude of the signed gap. 

As has been observed with complementary gaps over $d$ \cite{Gidney2024YokedSurfaceCodes, Toshio2025DecoderSwitching}, with increasing $\eta$, correlated decoders become more confident in a $Z$-memory (see Fig. \ref{fig:gap_over_eta}). The same effect does not emerge for $X$-memories for $CSS$ codes due to the nature of the logical. For $ZXXZ\square$ deformed codes, the $X$ and $Z$ memories are more confident with increasing $\eta$. Across $X$ and $Z$ memories, for high $\eta$, we find that the $ZXXZ\square$ deformed codes have more confident MWPM, while $CSS$ deformed codes have more confident \texttt{PyMatching} correlated decoding. At low $\eta$, we find the inverse to be the case. At moderate $\eta$, \texttt{PyMatching} correlated decoding is the most confident for all code types. We note that the decoder confidence does not match decoder performance with $\eta$, which can be seen by comparing directly to Fig. \ref{fig:deltacorr_vs_bias}.

For an $X$ or $Z$ memory experiment, we calculate the complementary gap, the signed complementary gap, and the conditional gap given success probability. To obtain these corrections, we add two additional boundary nodes to the matching graph. These nodes must align with the $Z_L$ or $X_L$ operator. We then fix one node as a boundary node. The other node remains a variable node. Both nodes are connected to the $X(Z)$ stabilizers for the $X(Z)$-memory to measure the $Z_L(X_L)$ gap. We complete one round of MWPM with the variable nodes off, and one round with the variable nodes fired. The variable node turned on represents the $X(Z)_L$ class, since a matching must include the boundary. MWPM picks the minimum weight path, independent of the logical class. If the minimum weight path coincides with whether a logical observable was flipped, the first round of decoding was successful. The complementary gap is the absolute difference in weight between the minimum weight path and the complementary path, $g = |W_{comp} - W_{min}|$. We find this by taking the difference in the path results from our two matching graphs, after noting which of the two matching MWPM chose. The signed gap is negative if the difference of the minimum weight path and the complementary path is negative. In decoding, we use the unsigned gap $g$ which is always positive.

\section{$CSS$ Correlated Decoding Algorithm}\label{sec:appendixd_decoding_algs}
The algorithm to implement the code-capacity correlated decoder is given in algorithm \ref{alg:correlated_decoder_vector}. We follow the algorithm in \cite{Delfosse2014CorrDecoding}, adjusting for noise bias in the erasure decoding step. 
\begin{algorithm}[h]
\caption{$CSS$ Correlated Decoding.}
\label{alg:correlated_decoder_vector}
\small
\begin{algorithmic}[1]
\REQUIRE Parity check matrix $\mathcal{H}_x, \mathcal{H}_z$, shots $N_{\text{shots}}$, physical error probability $p$, noise bias $\eta$, decoding order ($XZ$ or $ZX$)
\ENSURE Number of logical failures $\mathbf{N}_{fail}$

\STATE Generate the errors on qubits$\mathrm{E}$ biased Pauli channel on all qubits for $N_{\text{shots}}$
\STATE Decode on the matchgraph $\mathcal{M}_x(\mathcal{M}_z)$ generated from $\mathcal{H}_x(\mathcal{H}_z)$ in parallel for all $N_{\text{shots}}$
\STATE Calculate the correction $\mathcal{C}$ in $Z$($X$) for each shot

\FOR{$i = 1$ \TO $N_{\text{shots}}$}
    \STATE Set the conditional probability $P_c$ to $P(X|Z)$(or $P(Z|X)$) from equation \ref{eq:cond_probs}. 
    \STATE Update the weights $\mathbf{W}_i$ of $\mathcal{M}_z(\mathcal{M}_x)$ using the correction $\mathcal{C}_i$ with $w = \log{\frac{1-P_c}{P_c}}$ 
    \STATE Decode on the updated $\mathcal{M}_z(\mathcal{M}_x)$
\ENDFOR

\STATE Add together the logical errors divided by $N_{\text{shots}}$ for the first and second round of corrections to find $\mathbf{L}$.

\RETURN $\mathbf{L}$
\end{algorithmic}
\end{algorithm}
 
\section{Threshold Tables}
\label{sec:appendixe_threshold_tables}

We tabulate all of our thresholds. The thresholds are calculated with $1 \times 10^6$ shots, $d = 11,13,15,17,19$. All simulation results and scripts used to generate the data in this text are available at \cite{correlated_decoding_cd_compass_codes_2026}. All code is available in \cite{meinking_code_2026}.

\FloatBarrier
\providecommand{\noopsort}[1]{}\providecommand{\singleletter}[1]{#1}%

\end{document}